\pdfoutput=1
\documentclass[10pt,twocolumn]{article}
\usepackage[letterpaper,margin=0.75in,columnsep=0.25in]{geometry}
\usepackage[T1]{fontenc}
\usepackage{lmodern}
\usepackage{mathptmx}
\usepackage{amsmath,amssymb,amsthm}
\usepackage{graphicx}
\usepackage{booktabs}
\usepackage{threeparttable}
\usepackage{xcolor}
\usepackage{microtype}
\usepackage[font=small,labelfont=bf]{caption}
\usepackage[numbers,super,sort&compress]{natbib}
\setcitestyle{super,open={[},close={]},citesep={,}}
\usepackage[hidelinks]{hyperref}
\hypersetup{
  pdftitle={Beyond the Mean Gap: What Transferred-Angle QAOA Retains on a Superconducting Processor},
  pdfauthor={Sanim Yousuf Fahim and Mohammad Arif Hossain},
  pdfkeywords={Quantum approximate optimization algorithm, superconducting quantum processor, noise, error suppression, benchmarking}
}

\graphicspath{{figures/}{./}}

\newtheorem{proposition}{Proposition}

\newcommand{\gu}{g_{\mathrm{unif}}}
\newcommand{\gbar}{\bar g}

\usepackage{fancyhdr}
\fancypagestyle{preprint}{\fancyhf{}\fancyhead[L]{\footnotesize\itshape Preprint}\fancyhead[R]{\footnotesize\itshape Fahim and Hossain: Beyond the Mean Gap}\fancyfoot[C]{\thepage}}
\title{\bfseries Beyond the Mean Gap: What Transferred-Angle QAOA Retains on a Superconducting Processor}
\author{%
Sanim Yousuf Fahim$^{1}$ \quad and \quad Mohammad Arif Hossain$^{2,*}$\\[6pt]
\normalsize $^{1}$Daffodil International University, Dhaka, Bangladesh\\
\normalsize $^{2}$Department of Engineering Technology and Quantum Research Interdisciplinary Science and Education (QRISE) Center,\\
\normalsize Middle Tennessee State University, Murfreesboro, TN 37132, USA\\[3pt]
\normalsize $^{*}$Corresponding author: \href{mailto:mohammad.hossain@mtsu.edu}{mohammad.hossain@mtsu.edu}}
\date{\small\textit{Preprint}. This manuscript has not been peer reviewed.}

\begin{document}

\twocolumn[
\begin{@twocolumnfalse}
\maketitle
\thispagestyle{preprint}
\begin{abstract}
\noindent Hardware executions of the quantum approximate optimization algorithm (QAOA) are often validated by comparing the device's mean cost with a simulation, a check that can miss what matters for optimization. Under global depolarization, the mean normalized gap and the floor-corrected probability of every set of bitstrings, the optimum included, retain the same fraction of their noiseless improvement over uniform sampling. We test this prediction with depth-2, transferred-angle QAOA circuits for binary subproblems of an alternating-direction decomposition on the IBM Heron processor \texttt{ibm\_marrakesh}, with and without dynamical decoupling and twirling. At 12 qubits, the main-family circuits retained 29\% to 46\% of their mean-gap improvement but only 5\% to 15\% of their probability of the optimum, lower on at least nine of ten instances per run. Near-optimal bitstrings were lost about as much as the optimum, whereas the avoidance of poor bitstrings was retained best. In the first submission, a device-matched noise model reproduced the pattern but overestimated the retention of the optimum five- to eightfold. On a rugged planted family, the device found the optimum in 62\% of 768-shot blocks (noiseless: 98\%), within the range of local heuristics. Validations of optimization circuits should report retention of the optimum and of near-optimal sets.

\end{abstract}
\vspace{1.5em}
\end{@twocolumnfalse}
]

\section{Introduction}\label{sec:intro}

The quantum approximate optimization algorithm (QAOA)\cite{farhi2014quantum} prepares a parameterized state whose measurement outcomes are candidate solutions of a combinatorial problem. It has been executed on superconducting\cite{harrigan2021quantum,weidenfeller2022scaling,pelofske2024scaling} and trapped-ion\cite{pagano2020quantum} processors, and noise is the main obstacle to using it at scale.\cite{preskill2018nisq,stilckfranca2021limitations} Hardware runs are commonly summarized by an average quality of the sampled bitstrings, such as the mean cost or approximation ratio, and compared with a noiseless or noisy simulation,\cite{harrigan2021quantum,weidenfeller2022scaling,lubinski2024optimization} and how to benchmark quantum optimization fairly has become a central question for the field.\cite{abbas2024challenges,koch2026quantum} For optimization, however, what counts is whether the device returns good solutions at all: the time to solution of a sampler depends on the probability it places on the optimum,\cite{ronnow2014defining,albash2018demonstration,lotshaw2022scaling} and a mean can hide where the probability mass went.

This work asks whether the mean-gap agreement of a hardware run with its simulation predicts how much of that probability survives. We study QAOA circuits that arise as the binary subproblem of the alternating direction method of multipliers (ADMM)\cite{boyd2011distributed} for mixed-binary problems, a decomposition in which a quantum heuristic can be delegated the quadratic unconstrained binary optimization (QUBO) step at every outer iteration.\cite{gambella2020multiblock} The circuits use transferred angles: depth-2 angles trained once on small held-out instances and reused,\cite{brandao2018forfixed,akshay2021concentration,shaydulin2023parameter} in the spirit of fixed-schedule protocols,\cite{montanezbarrera2025toward} so that every shot is a candidate solution and no optimizer loop runs on the device. In noiseless simulation these circuits concentrate probability on the optimum strongly enough to solve the 12-variable main-family subproblems within a few hundred shots, which makes them a demanding test of what the hardware preserves.

The paper makes four contributions.
\begin{enumerate}
\item \emph{What a mean-gap validation can detect} (Section~\ref{sec:prop}). Under global depolarization, the mean normalized gap and the floor-corrected probability of every set of bitstrings retain the same fraction of their noiseless improvement over uniform sampling. A comparison of mean gaps therefore resolves fidelity only in proportion to the distance of the ideal output from uniform, is ill-conditioned for near-uniform outputs, and predicts that the optimum is retained as well as the mean gap.
\item \emph{An 8-qubit validation} (Section~\ref{subsec:n8}). On five instances in two sessions, mean-gap agreement carried almost no information for random angles, whose outputs are close to uniform. For transferred angles, the device-matched noise model retained 90\% of the noiseless improvement over uniform sampling, the hardware 68\% to 73\% in one session and 49\% to 57\% in a second session about ten hours later, so the model was optimistic on every instance and the change between sessions was comparable to the model's error.
\item \emph{The probability of the optimum at 12 qubits} (Sections~\ref{subsec:n12}--\ref{subsec:threshold}). In every main-family run, the floor-corrected retention of the probability of the optimum was three to eight times lower than the mean-gap retention, contrary to the global-depolarization prediction: the runs placed a median of 5.5 to 14 shots of 4096 on the optimum per instance, where global depolarization at the measured mean-gap retention predicts 23 to 33. Threshold-resolved retention shows that near-optimal bitstrings were lost about as much as the optimum, whereas the avoidance of poor bitstrings was retained best, so the mean gap mostly measures the latter. A device-matched noise model reproduces the pattern but not its size.
\item \emph{Consequences for optimization} (Sections~\ref{subsec:planted}--\ref{subsec:admm}). The hardware time to solution was 5 to 14 times the noiseless value and, in the first submission, longer than that of three tuned classical heuristics on every instance. On a rugged planted family, the advantage of transferred QAOA over local heuristics in noiseless simulation did not survive execution. Twirling lowered and dynamical decoupling raised the retention, and two small studies with the ADMM binary step on the device illustrate how hardware errors propagate to the decomposition.
\end{enumerate}

\begin{figure*}[t]
\centering
\includegraphics[width=\textwidth]{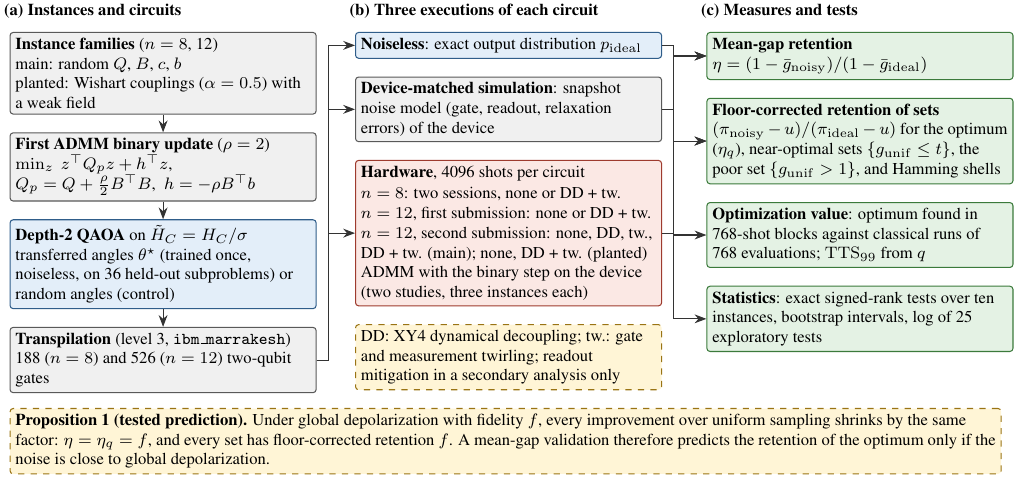}
\caption{Study design. (a) Instances, the first ADMM binary subproblem, depth-2 QAOA circuits with transferred or random angles, and their transpilation for the device. (b) Each circuit is evaluated exactly, in a device-matched noisy simulation, and on hardware under several suppression settings. (c) Retention measures, optimization value, and tests. Proposition~\ref{prop:depol} predicts that under global depolarization the mean-gap retention $\eta$ equals the retention $\eta_q$ of the optimum and of every other set.}\label{fig:design}
\end{figure*}

\section{Related Work}\label{sec:related}

\emph{QAOA on hardware.} QAOA has been executed on superconducting\cite{harrigan2021quantum,weidenfeller2022scaling,pelofske2024scaling} and trapped-ion\cite{pagano2020quantum} processors. These studies characterize performance by the mean energy or approximation ratio of the sampled output as a function of problem size and depth and compare it with noiseless or noisy simulation, and application-oriented benchmark suites weigh result quality against execution cost.\cite{lubinski2024optimization} We ask instead how much of the probability of the optimum, which sets the time to solution, survives on hardware, and whether the mean-based comparison predicts it.

\emph{Noise in QAOA.} Numerical and analytical studies of local noise in QAOA circuits\cite{xue2021effects,marshall2020characterizing} find that local depolarizing noise reduces the output approximately as a global fidelity factor.\cite{marshall2020characterizing} Noise also flattens the cost landscape,\cite{wang2021noise} limits the depth at which optimization on noisy devices can outperform efficient classical algorithms,\cite{stilckfranca2021limitations} and makes the number of measurements, and hence the time to solution, grow with gate error.\cite{lotshaw2022scaling}

\emph{Fidelity estimation under white noise.} Fidelity estimates from sampled outputs, such as the linear cross-entropy benchmark,\cite{arute2019quantum} rely on describing the noisy output as a mixture of the ideal output and the uniform distribution, and for random circuits local noise is indeed scrambled into global white noise.\cite{dalzell2021random} QAOA circuits are structured and concentrate probability on few bitstrings; we test whether the mixture model holds for the quantity that matters for optimization.

\emph{Error suppression and mitigation.} Dynamical decoupling\cite{ezzell2023dynamical} and twirling\cite{wallman2016noise} act at the circuit level; zero-noise extrapolation\cite{temme2017error,kandala2019error} and readout correction\cite{bravyi2021mitigating} correct expectation values and have been applied at the scale of more than one hundred qubits.\cite{kim2023evidence} Corrected expectation values do not restore the samples that an optimization heuristic returns, so we evaluate suppression settings and use readout mitigation only as a check.

\section{Problem, Circuits, and Retention Measures}\label{sec:setup}

\subsection{Binary Subproblems of ADMM}\label{subsec:problem}

We consider mixed-binary problems of the form
\begin{equation}
\min_{x\in\mathbb{R}^n,\,z\in\{0,1\}^n}\ \tfrac12\|x\|_2^2-c^{\top}x+z^{\top}Qz\quad\text{subject to}\quad x+Bz=b,
\label{eq:mbo}
\end{equation}
with $Q,B\in\mathbb{R}^{n\times n}$, $Q$ symmetric, and $c,b\in\mathbb{R}^n$, an instance of the quadratic-plus-convex class studied by Gambella and Simonetto.\cite{gambella2020multiblock} A two-block ADMM with penalty $\rho$ alternates a binary update, a continuous update, and a dual update.\cite{boyd2011distributed} Starting from $x^0=\lambda^0=0$, the binary update of the first iteration is the QUBO
\begin{equation}
\begin{gathered}
\min_{z\in\{0,1\}^n}\ C(z)=z^{\top}Q_pz+h^{\top}z,\\
Q_p=Q+\tfrac{\rho}{2}B^{\top}B,\qquad h=-\rho B^{\top}b,
\end{gathered}
\label{eq:qubo}
\end{equation}
which is the circuit input of this study. All circuits use $\rho=2$. The \emph{main family} draws the entries of $Q$ uniformly from $[-1,1]$ (symmetrized), those of $B$ from $[-0.7,0.7]$, and those of $c$ and $b$ from $[-1,1]$. The \emph{planted family} takes $Q$ from the Wishart planted ensemble:\cite{hamze2020wishart} Ising couplings $J=-WW^{\top}/n$ with zero diagonal, where the $\alpha n$ Gaussian columns of $W$ ($\alpha=0.5$) are orthogonal to a planted configuration hidden by a random gauge, mapped to a QUBO scaled to unit maximum magnitude and weakly coupled to the continuous block ($B$ from $[-0.05,0.05]$, $c$ and $b$ from $[-0.1,0.1]$). Its landscape is rugged, and its weak linear terms leave the complement of the optimum close to it in cost. Instances are generated from fixed seeds (main family: 0 to 9; planted family: 30000 to 30009), and exhaustive enumeration of the $2^n$ bitstrings gives the optimum $C^*$ and the mean cost $\bar C$ of every subproblem.

\subsection{QAOA Circuits and Transferred Angles}\label{subsec:circuits}

With $z_i=(1-s_i)/2$ and $s_i\mapsto Z_i$, the cost \eqref{eq:qubo} defines a diagonal cost Hamiltonian $H_C$. QAOA of depth $p$ prepares
\begin{equation}
|\boldsymbol\gamma,\boldsymbol\beta\rangle=\prod_{l=1}^{p}e^{-i\beta_l H_M}e^{-i\gamma_l\tilde H_C}\,|+\rangle^{\otimes n},\qquad H_M=\sum_iX_i,
\label{eq:qaoa}
\end{equation}
with the factors ordered from $l=1$ on the right, and $\tilde H_C=H_C/\sigma$, where $\sigma$ is the root-mean-square of the non-identity Pauli coefficients of $H_C$; the normalization makes angle scales comparable across instances.\cite{sureshbabu2024parameter} We use $p=2$. \emph{Transferred angles} are fixed in advance: for each family, they minimize the mean exact expected normalized gap (defined below) over 36 held-out subproblems, the first, third, and fifth of the exact-solver ADMM trajectories of 12 held-out instances, with Nelder--Mead\cite{nelder1965simplex} from one linear-ramp and five random starting points. The resulting angles are $(\beta_1,\beta_2,\gamma_1,\gamma_2)=(0.672,0.365,-0.126,-0.253)$ for the main family and $(-0.228,0.163,0.100,-0.301)$ for the planted family. \emph{Random angles}, used as a control, are drawn uniformly from $[-\pi/4,\pi/4]$ with a fixed seed per instance. The 8-qubit study applied them to the unnormalized Hamiltonian and the 12-qubit study to the normalized one; because every random-angle output is compared only with the noiseless output of the same circuit, this difference changes which outputs the control produces, not how retention is measured.

\subsection{Normalized Gap and Retention}\label{subsec:measures}

The normalized gap of a bitstring is
\begin{equation}
\gu(z)=\frac{C(z)-C^*}{\bar C-C^*},
\label{eq:gu}
\end{equation}
which is 0 at the optimum and 1 for the expected cost of a uniform guess. For an output distribution, $\gbar$ denotes its mean normalized gap; the uniform distribution has $\gbar=1$. For a circuit with noiseless mean gap $\gbar_{\mathrm{ideal}}$ and measured mean gap $\gbar_{\mathrm{noisy}}$, the \emph{mean-gap retention}
\begin{equation}
\eta=\frac{1-\gbar_{\mathrm{noisy}}}{1-\gbar_{\mathrm{ideal}}}
\label{eq:eta}
\end{equation}
is the fraction of the noiseless improvement over uniform sampling that survives. For a set $\Omega$ of bitstrings with uniform share $u=|\Omega|/2^n$, noiseless probability $\pi_{\mathrm{ideal}}$, and measured probability $\pi_{\mathrm{noisy}}$, the \emph{floor-corrected retention} of $\Omega$ is $(\pi_{\mathrm{noisy}}-u)/(\pi_{\mathrm{ideal}}-u)$. For the optimum, with probability $q$ and $u=2^{-n}$ for a unique optimum, it is
\begin{equation}
\eta_q=\frac{q_{\mathrm{hw}}-2^{-n}}{q_{\mathrm{ideal}}-2^{-n}}.
\label{eq:etaq}
\end{equation}
Every subproblem in this study has a unique optimum. With $N=4096$ shots at $n=12$, a single optimal shot equals the floor $2^{-n}$, so a run with one optimal shot has $\eta_q=0$.

\section{What a Mean-Gap Validation Can Detect}\label{sec:prop}

\begin{proposition}[Global depolarization shrinks every improvement equally]\label{prop:depol}
Suppose the device returns a sample of the ideal output distribution with probability $f$ and a uniformly random bitstring otherwise. Then
\begin{equation}
1-\gbar_{\mathrm{noisy}}=f\,\bigl(1-\gbar_{\mathrm{ideal}}\bigr).
\label{eq:depol}
\end{equation}
Hence (a) when $\gbar_{\mathrm{ideal}}\neq1$, the mean-gap retention $\eta$ equals $f$; (b) a change $\Delta f$ in fidelity moves $\gbar_{\mathrm{noisy}}$ by only $|1-\gbar_{\mathrm{ideal}}|\,|\Delta f|$; (c) the same holds for averages over instances that share the same $f$; and (d) for any set $\Omega$ with uniform share $u$, $\pi_{\mathrm{noisy}}-u=f\,(\pi_{\mathrm{ideal}}-u)$, so the floor-corrected retention of every set with $\pi_{\mathrm{ideal}}\neq u$, the optimum included, also equals $f$.
\end{proposition}

\begin{proof}
The output distribution is $f\,p_{\mathrm{ideal}}+(1-f)\,p_{\mathrm{unif}}$. The mean gap is the expectation of $\gu$ under the output distribution and is therefore linear in it, and the uniform distribution has mean gap 1, so $\gbar_{\mathrm{noisy}}=f\,\gbar_{\mathrm{ideal}}+1-f$, which rearranges to \eqref{eq:depol}. Dividing by $1-\gbar_{\mathrm{ideal}}\neq0$ gives (a); changing $f$ by $\Delta f$ changes the right side of \eqref{eq:depol} by $(1-\gbar_{\mathrm{ideal}})\,\Delta f$, which gives (b); averaging \eqref{eq:depol} over instances with the same $f$ gives (c). The mixture places probability $f\,\pi_{\mathrm{ideal}}+(1-f)\,u$ on $\Omega$, which rearranges to (d).
\end{proof}

The mixture argument is the one that underlies fidelity estimation from sampled outputs;\cite{arute2019quantum} we use it to derive predictions that can be tested on optimization circuits. A validation that compares hardware and model through mean gaps can therefore resolve a fidelity difference only in proportion to $|1-\gbar_{\mathrm{ideal}}|$. Because noise moves every mean gap toward 1, a circuit whose ideal mean gap exceeds 1, that is, whose angles are anti-aligned with the cost, can make a noisier device appear better. Under local depolarizing noise, the output of a QAOA circuit can be approximated by a global depolarizing channel with an effective fidelity,\cite{marshall2020characterizing} but real noise is not globally depolarizing. We therefore use $\eta$ as an effective fidelity for comparing a noise model with the device on the same circuits, not as an estimate of circuit fidelity, and test three predictions. If instances executed with the same circuit structure share an effective fidelity, an assumption that Proposition~\ref{prop:depol} does not itself make, $\eta$ should be similar across them; by Proposition~\ref{prop:depol}(b), it should be poorly determined when $|1-\gbar_{\mathrm{ideal}}|$ is small; and by Proposition~\ref{prop:depol}(d), it should equal the floor-corrected retention of every good set, including the optimum on which the time to solution depends. On hardware, the transferred-angle circuits are consistent with the first prediction, but the random-angle circuits at $n=12$ are not, which shows that the effective fidelity depends on the angles as well as on the circuit structure; the second prediction holds; and the third fails on every main-family run.

\section{Results}\label{sec:results}

\subsection{An 8-Qubit Validation}\label{subsec:n8}

\begin{figure*}[t]
\centering
\includegraphics[width=\textwidth]{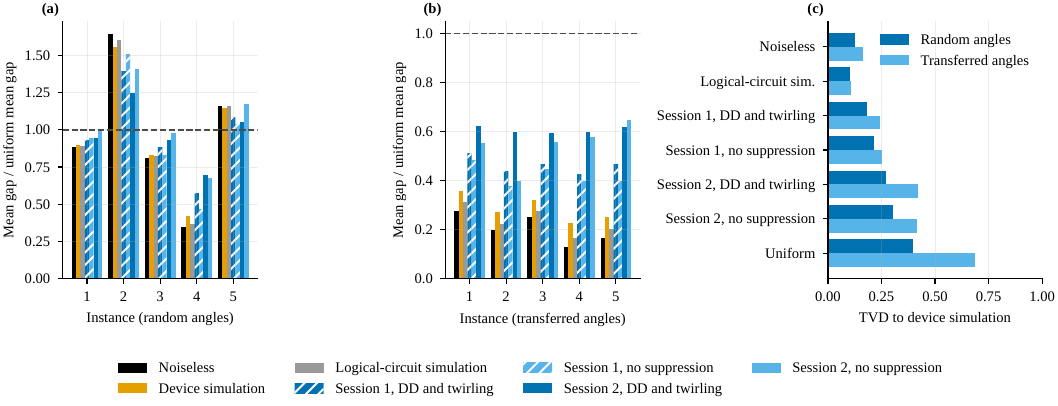}
\caption{Hardware at $n=8$ on \texttt{ibm\_marrakesh}, both sessions. Mean gap relative to uniform for (a) random and (b) transferred angles, per instance; the dashed line marks the uniform distribution. (c) Total variation distance (TVD) to the device-matched simulation, mean over instances.}\label{fig:n8}
\end{figure*}

Figure~\ref{fig:n8} compares each hardware distribution with the noiseless distribution at the same angles, the device-matched and logical-circuit simulations, and the uniform distribution. With random angles, the per-instance noiseless mean gap ranges from about 0.35 to 1.65 of the uniform value, but these biases point in arbitrary directions: the means over instances of the noiseless, simulated, and hardware distributions all lie between 0.955 and 1.045 of it, so by Proposition~\ref{prop:depol}(c) their agreement carries almost no information. On the two instances whose random angles are anti-aligned with the cost ($\gbar_{\mathrm{ideal}}>1$), noise pulls the output toward uniform and improved the gap in all but one hardware run, so a simulation with less noise than the executed circuit appears to underperform the hardware. This mechanism produced a spurious hardware-versus-model discrepancy in our initial implementation, which simulated the unrouted circuit.

With transferred angles, the noiseless output is far from uniform, and the comparison is informative. The device-matched simulation retained 90\% of the noiseless improvement over uniform sampling (96\% when the same noise acts on the unrouted logical circuit), whereas hardware retained 68\% and 73\% in the first session and 49\% and 57\% in the second, and the total variation distance to the simulated distribution rose from 0.16 (noiseless) to between 0.24 and 0.42 (hardware), against 0.69 for the uniform distribution. The noise model was therefore optimistic on every instance, and the change between sessions was comparable to the model's error. Hardware output nonetheless stayed well below uniform on every instance in both sessions (0.38 to 0.65 of the uniform mean gap), so the transferred-angle state survived execution, although heavily degraded.

At this size, the first two predictions of Section~\ref{sec:prop} hold per instance. Under the device-matched simulation, $\eta$ lies between 0.89 and 0.91 on all five transferred-angle instances and between 0.86 and 0.91 on all five random-angle instances. On hardware, $\eta$ for transferred angles ranges from 0.64 to 0.78 in the first session and from 0.42 to 0.75 in the second. For random angles, it ranges from 0.60 to 0.82 and from 0.39 to 0.63 on the two instances with $|1-\gbar_{\mathrm{ideal}}|>0.6$, close to the transferred-angle values of the same session, and scatters from 0.23 to 0.92 and from 0.03 to 1.06 on the three instances with $|1-\gbar_{\mathrm{ideal}}|<0.2$, where Proposition~\ref{prop:depol}(b) predicts an ill-conditioned estimate. The 8-qubit study measured distributions, not whether hardware samples solve subproblems, which the 12-qubit submissions address.

\subsection{Success and Mean-Gap Retention at 12 Qubits}\label{subsec:n12}

\begin{figure*}[t]
\centering
\includegraphics[width=\textwidth]{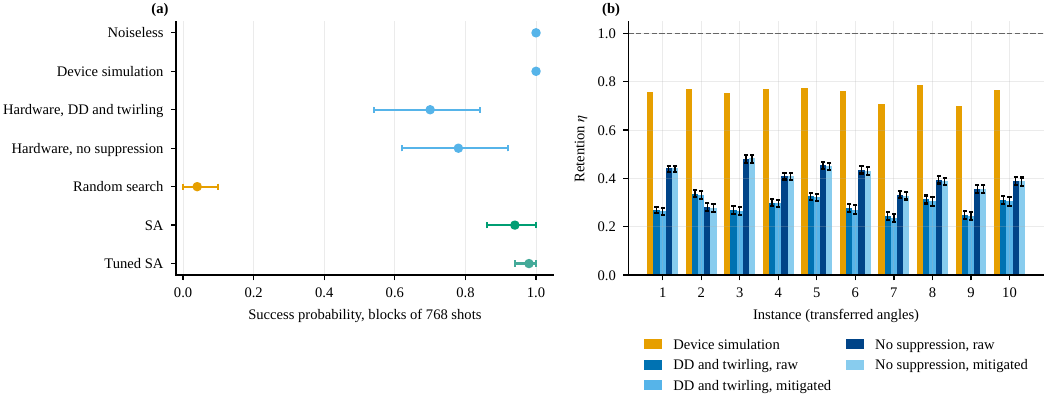}
\caption{First 12-qubit submission (ten main-family instances, transferred angles). (a) Fraction of 768-shot blocks containing the subproblem optimum, with random search, SA, and tuned SA at 768 evaluations, mean and 95\% interval. (b) Mean-gap retention $\eta$ per instance for the device-matched simulation and for hardware with and without suppression (dynamical decoupling, DD, and twirling together), raw and readout-mitigated, with 95\% shot-bootstrap intervals.}\label{fig:n12first}
\end{figure*}

Table~\ref{tab:n12} summarizes the 12-qubit submissions. In the first, transferred-angle circuits found the optimum in 70\% (with suppression) and 78\% (without) of 768-shot blocks, against 100\% in noiseless and device-matched simulation and 4\% for random search, and beat random search on all ten instances in both modes ($P=0.002$) (Figure~\ref{fig:n12first}). Simulated annealing (SA) and SA with tuned parameters (tuned SA; Section~\ref{subsec:scoring}) found it in 94\% and 98\% of their runs, more often than the hardware ($P=0.016$ with suppression and $P=0.063$ without, against each; neither below the Bonferroni threshold). The mean retention $\eta$ was 0.29 with and 0.40 without suppression, against 0.75 for the device-matched simulation, whose prediction was optimistic on every instance and whose per-instance spread (0.70 to 0.78) was far smaller than the gap between prediction and hardware. Readout mitigation changed the transferred-angle $\eta$ by less than 0.01, so readout error is not what the model misses. For random angles, $\eta$ ranged from $-0.49$ to 10.6 on the five instances with near-uniform ideal output ($|1-\gbar_{\mathrm{ideal}}|\le0.053$), the ill-conditioning that Proposition~\ref{prop:depol}(b) predicts. On the other five ($|1-\gbar_{\mathrm{ideal}}|$ from 0.26 to 0.65), where $\eta$ is well determined, it ranged from 0.11 to 0.27 with suppression and from $-0.09$ to 0.81 without, against 0.24 to 0.34 and 0.28 to 0.48 for transferred angles in the same runs and 0.53 to 0.75 in the device-matched simulation. These circuits have the same 526 two-qubit gates as the transferred-angle circuits and a similar depth, so the effective fidelity of the device depended on the angles, consistent with errors whose effect depends on the rotation angles, such as coherent errors, which we did not identify.

\begin{table*}[t]
\begin{threeparttable}
\caption{Hardware at $n=12$ ($p=2$, transferred angles, ten instances per family, first ADMM subproblem).}\label{tab:n12}
\begin{tabular*}{\textwidth}{@{\extracolsep\fill}lcccc}
\toprule
Condition & $\Pr_{768}$ & $\eta$ & $q$ & $\mathrm{TTS}_{99}$ \\
\midrule
\multicolumn{5}{@{}l}{\emph{Main family}} \\
Noiseless & 1.00 & 1 & 0.0185 & 246 \\
Device model (first)$^{\rm a)}$ & 1.00 & 0.75 (0.70--0.78) & 0.0072 & 646 \\
First, DD + tw. & 0.70 [0.54, 0.84] & 0.29 (0.24--0.34) & 0.0013 & 3456 \\
First, none & 0.78 [0.62, 0.92] & 0.40 (0.28--0.48) & 0.0017 & 2692 \\
Second, none & 0.82 [0.66, 0.96] & 0.44 (0.38--0.50) & 0.0024 & 2071 \\
Second, DD & 0.90 [0.80, 1.00] & 0.46 (0.40--0.53) & 0.0034 & 1352 \\
Second, tw. & 0.76 [0.66, 0.86] & 0.38 (0.32--0.41) & 0.0022 & 2094 \\
Second, DD + tw. & 0.78 [0.66, 0.88] & 0.37 (0.31--0.42) & 0.0016 & 2917 \\
Classical$^{\rm b)}$ & 0.94--0.98 & n/a & n/a & 101--492 \\
\midrule
\multicolumn{5}{@{}l}{\emph{Planted family}} \\
Noiseless & 0.98 & 1 & 0.0049 & 942 \\
Second, none & 0.62 [0.48, 0.74] & 0.36 (0.29--0.40) & 0.0011 & 4242 \\
Second, DD + tw. & 0.40 [0.28, 0.52] & 0.23 (0.20--0.27) & 0.0007 & 6285 \\
Classical$^{\rm c)}$ & 0.54--0.82 & n/a & n/a & n/a \\
\bottomrule
\end{tabular*}
\begin{tablenotes}
\item $\Pr_{768}$: fraction of 768-shot blocks (or 768-evaluation runs) containing the optimum, mean over instances with 95\% interval (noiseless: exact value $1-(1-q)^{768}$); $\eta$: mean-gap retention, mean (range over instances); $q$: probability of the optimum; $\mathrm{TTS}_{99}$: evaluations to find the optimum with probability 0.99; $q$ and $\mathrm{TTS}_{99}$ are medians over instances. Unmitigated counts. First, second: submission; DD: XY4 dynamical decoupling; tw.: gate and measurement twirling.
\item[a)\ ] $q$ and $\mathrm{TTS}_{99}$ from the re-execution of the device-matched simulation (post hoc).
\item[b)\ ] First-submission instances. $\Pr_{768}$: SA and tuned SA (random search: 0.04). $\mathrm{TTS}_{99}$: SDP+LS 101, tabu search with restarts 192, SA 365, local search 384, PT 492. SDP+LS: semidefinite-programming rounding followed by local search; PT: parallel tempering.
\item[c)\ ] Tuned SA and PT 0.54, local search 0.62, tabu search with restarts 0.78 [0.66, 0.90], SDP+LS 0.82 [0.56, 1.00] (random search: 0.20).
\end{tablenotes}
\end{threeparttable}
\end{table*}

\subsection{The Probability of the Optimum}\label{subsec:optimum}

\begin{figure*}[t]
\centering
\includegraphics[width=\textwidth]{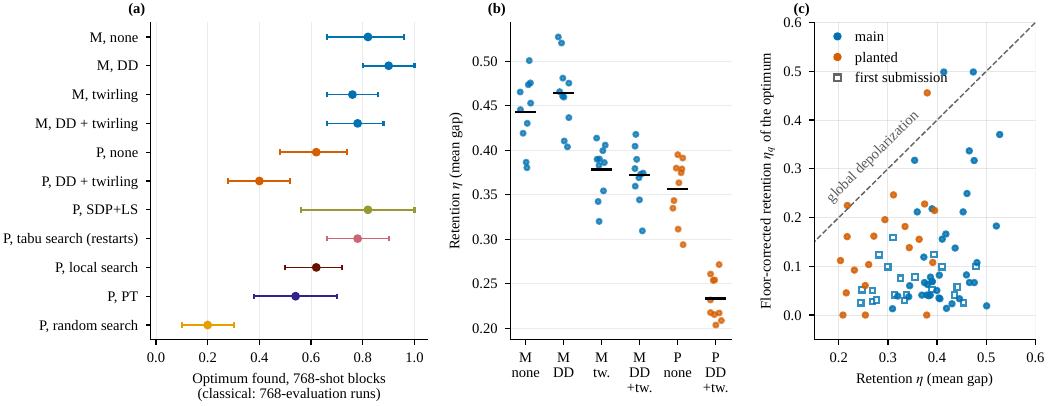}
\caption{Second 12-qubit submission (ten instances per family; M: main, P: planted). (a) Fraction of 768-shot blocks containing the subproblem optimum per suppression mode, and of 768-evaluation runs of SDP+LS, tabu search with restarts, local search, PT, and random search on the planted subproblems; mean and 95\% interval. (b) Mean-gap retention $\eta$ per instance (horizontal lines: means). (c) Floor-corrected retention $\eta_q$ of the probability of the optimum against $\eta$, per instance and run (filled: second submission; open squares: first submission, main family); the dashed line is the global-depolarization prediction of Proposition~\ref{prop:depol}(d).}\label{fig:n12second}
\end{figure*}

Hardware violates the global-depolarization prediction of Proposition~\ref{prop:depol}(d). On every main-family mode of both submissions, the floor-corrected retention $\eta_q$ of the probability of the optimum was below the mean-gap retention $\eta$ of the same run on at least nine of ten instances (Table~\ref{tab:etaq}; signed-rank $P\le0.006$ in every mode), by a median factor of 3 to 8 per instance; on the planted circuits, the factor was 1.8 to 2.5 (Figure~\ref{fig:n12second}c). The medians of $\eta_q$ on the main-family runs were 0.046 to 0.147, rested on 5.5 to 14 optimal shots of 4096 per instance, and had wide intervals, but the ordering $\eta_q<\eta$ held per instance, and every 95\% interval of a median $\eta_q$ lies below the median $\eta$ of the same run (Table~\ref{tab:etaq}). The ordering is not an artifact of few counts. Under global depolarization at the measured $\eta$, an instance would place $4096\,[\eta\,q_{\mathrm{ideal}}+(1-\eta)\,2^{-n}]$ shots on the optimum in expectation, a median of 23 to 33 per main-family run against the observed 5.5 to 14 (Table~\ref{tab:etaq}); the observed count fell below this expectation on at least nine of ten instances of every run, and an exact one-sided binomial test rejected the expectation at the 0.05 level on 7 to 10 of the ten instances of each main-family run (post hoc). The device-matched model, with a median of 29.5 optimal shots per instance, shows the ordering as well. On the planted circuits, the expected counts were only 8.1 and 5.6 (medians) against the observed 4.5 and 3, and the binomial test rejected the expectation on 2 of ten instances per run, so the planted-family factor rests on few counts and is less certain than the ordering on the main family.

The device-matched noise model of the first submission shows the same pattern at a higher level (median $\eta_q$ of 0.365 against median $\eta$ of 0.763, lower on nine of ten instances, signed-rank $P=0.006$), so part of the effect is generic to local noise of the kind the model contains. The model misses the size of the effect: on the first submission, the hardware's $\eta_q$ was lower than the model's by a factor of 5.4 to 8.0, and its $\eta$ by a factor of only 1.9 to 2.6.

\begin{table*}[t]
\begin{threeparttable}
\caption{Floor-corrected retention of the probability of the optimum against mean-gap retention ($n=12$, ten instances per row, unmitigated counts).}\label{tab:etaq}
\setlength{\tabcolsep}{4pt}
\begin{tabular*}{\textwidth}{@{\extracolsep\fill}llccccccc}
\toprule
Run & Mode & Counts & Exp. & $\eta_q$ [95\% CI] & $\eta$ & Ratio & Below & $P$ \\
\midrule
First, main & DD + tw. & 5.5 (2) & 22.8 & 0.046 [0.024, 0.082] & 0.289 & 6.4 & 10/10 & 0.002 \\
First, main & none & 7 (2) & 29.0 & 0.068 [0.032, 0.109] & 0.402 & 6.1 & 10/10 & 0.002 \\
Second, main & none & 10 (2) & 31.1 & 0.065 [0.023, 0.212] & 0.449 & 6.6 & 9/10 & 0.004 \\
Second, main & DD & 14 (5) & 32.7 & 0.147 [0.078, 0.237] & 0.463 & 3.0 & 10/10 & 0.002 \\
Second, main & tw. & 9 (3) & 27.5 & 0.046 [0.034, 0.208] & 0.388 & 8.1 & 9/10 & 0.006 \\
Second, main & DD + tw. & 6.5 (2) & 27.1 & 0.068 [0.038, 0.112] & 0.374 & 5.6 & 10/10 & 0.002 \\
Second, planted & none & 4.5 (1) & 8.1 & 0.189 [0.076, 0.290] & 0.369 & 1.8 & 9/10 & 0.006 \\
Second, planted & DD + tw. & 3 (1) & 5.6 & 0.098 [0.000, 0.165] & 0.225 & 2.5 & 9/10 & 0.004 \\
\midrule
Model (first)$^{\rm a)}$ & simulation & 29.5 (14) & 56.0 & 0.365 [0.282, 0.504] & 0.763 & 2.1 & 9/10 & 0.006 \\
\bottomrule
\end{tabular*}
\begin{tablenotes}
\item Counts: optimal bitstrings among 4096 shots, median (minimum) over instances; Exp.: optimal count expected under global depolarization at the measured $\eta$, median over instances (post hoc); $\eta_q$: median with 95\% confidence interval (CI; bootstrap over instances with binomial resampling of the counts); $\eta$: median over instances (Table~\ref{tab:n12} reports means); ratio: median over instances of $\eta/\eta_q$; below: instances with $\eta_q<\eta$; $P$: exact signed-rank test of $\eta_q$ against $\eta$. Global depolarization predicts $\eta_q=\eta$.
\item[a)\ ] Device-matched simulation of the first submission, re-executed (post hoc).
\end{tablenotes}
\end{threeparttable}
\end{table*}

\subsection{Where the Probability Is Lost}\label{subsec:threshold}

\begin{figure*}[t]
\centering
\includegraphics[width=\textwidth]{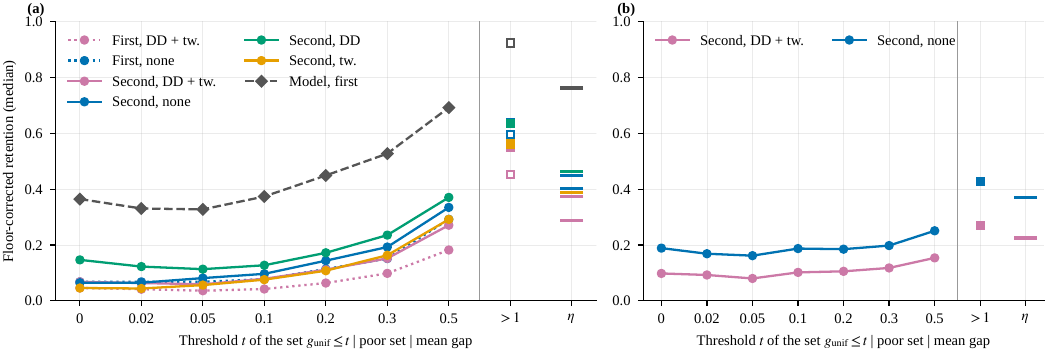}
\caption{Where the probability is lost ($n=12$, medians over ten instances, post hoc). Floor-corrected retention of the near-optimal sets $\{\gu\le t\}$ against the threshold $t$ ($t=0$: the optimum), of the ideal state's probability deficit on the poor set $\{\gu>1\}$ (squares; open: first submission), and the mean-gap retention $\eta$ of the same run (bars) for (a) the main family, including the device-matched model of the first submission, and (b) the planted family. Global depolarization predicts that every point of a run lies at its $\eta$.}\label{fig:threshold}
\end{figure*}

To locate the loss, we resolved the floor-corrected retention by threshold from the recorded counts and the exact noiseless distributions (Table~\ref{tab:threshold} and Figure~\ref{fig:threshold}; post hoc); global depolarization would give every set the same value. On every main-family run, the near-optimal sets $\{\gu\le t\}$ with $0<t\le0.1$, the best 0.2\% of bitstrings, were retained about as little as the optimum, and for every $t$ from 0.05 to 0.5 the retention was below $\eta$ on every instance; only broad sets approached $\eta$ ($t=0.5$, the best 11\% of bitstrings). The device best preserved the ideal state's avoidance of poor bitstrings: the probability deficit of $\{\gu>1\}$, almost half of all bitstrings, was retained more than $\eta$ on every instance of every run, planted circuits included. Nor did the probability lost from the optimum go mainly to its single-flip neighbors: on the main-family runs, the optimum together with its Hamming-distance-1 neighbors was retained at only 0.08 to 0.18, less than $\eta$ on every instance. On the planted circuits without suppression, the distance-1 shell alone kept more than $\eta$ (median 0.58), consistent with errors that flip single bits of a near-degenerate optimum, but the optimum itself was retained at 0.19. The mean-gap retention therefore measures mostly how well the device avoids poor solutions, not how well it keeps good ones.

The floor correction assumes that noise moves probability toward the uniform distribution. Relaxation instead drives qubits toward $|0\rangle$, which in our encoding lowers the number of ones in a bitstring. We checked this assumption on the counts (post hoc). Under global depolarization at the measured $\eta$, the mean number of ones per shot would be $\eta\,w_{\mathrm{ideal}}+(1-\eta)\,n/2$, where $w_{\mathrm{ideal}}$ is its noiseless value. On every main-family run, the observed mean was within 0.31 of this prediction on every instance (run medians from $-0.06$ to $+0.14$, out of 12 bits). On the planted circuits without suppression, it was lower on all ten instances, by 0.65 to 0.91 (medians 5.21 observed against 5.98 predicted; signed-rank $P=0.002$), of which readout asymmetry, estimated from the measured assignment errors, accounts for about 0.03; with dynamical decoupling and twirling the shift vanished (median 0.00). The floor-corrected values of that one run therefore include a bias toward $|0\cdots0\rangle$ that the correction does not model.

\begin{table*}[t]
\begin{threeparttable}
\caption{Threshold-resolved floor-corrected retention ($n=12$, medians over ten instances, post hoc).}\label{tab:threshold}
\begin{tabular*}{\textwidth}{@{\extracolsep\fill}llccccccc}
\toprule
 & & & \multicolumn{4}{c}{$\{\gu\le t\}$} & & \\
\cmidrule{4-7}
Run & Mode & $\eta$ & 0.05 & 0.1 & 0.2 & 0.5 & $\{\gu>1\}$ & $d\le1$ \\
\midrule
First, main & DD + tw. & 0.289 & 0.036 & 0.042 & 0.064 & 0.182 & 0.45 & 0.077 \\
First, main & none & 0.402 & 0.068 & 0.078 & 0.114 & 0.292 & 0.60 & 0.096 \\
Second, main & none & 0.449 & 0.081 & 0.096 & 0.144 & 0.335 & 0.64 & 0.136 \\
Second, main & DD & 0.463 & 0.113 & 0.127 & 0.172 & 0.370 & 0.64 & 0.178 \\
Second, main & tw. & 0.388 & 0.056 & 0.076 & 0.108 & 0.292 & 0.56 & 0.120 \\
Second, main & DD + tw. & 0.374 & 0.059 & 0.079 & 0.111 & 0.271 & 0.55 & 0.130 \\
Second, planted & none & 0.369 & 0.162 & 0.187 & 0.185 & 0.251 & 0.43 & 0.441 \\
Second, planted & DD + tw. & 0.225 & 0.080 & 0.102 & 0.105 & 0.154 & 0.27 & 0.161 \\
\midrule
Model (first) & simulation & 0.763 & 0.327 & 0.374 & 0.449 & 0.692 & 0.92 & 0.441 \\
\bottomrule
\end{tabular*}
\begin{tablenotes}
\item $\{\gu\le t\}$: bitstrings within normalized gap $t$ of the optimum; $\{\gu>1\}$: retention of the ideal state's probability deficit on the bitstrings worse than a uniform guess; $d\le1$: the optimum with its Hamming-distance-1 neighbors. Global depolarization predicts that every column equals $\eta$.
\end{tablenotes}
\end{threeparttable}
\end{table*}

\subsection{Time to Solution and the Planted Family}\label{subsec:planted}

The hardware time to solution, $\mathrm{TTS}_{99}$, the number of evaluations needed to find the optimum with probability 0.99, was 1352 to 3456 shots in the main-family runs (medians), 5 to 14 times the noiseless 246, and on the first submission tabu search with restarts, parallel tempering (PT), and SA needed fewer evaluations than the hardware on all ten instances ($P=0.002$ for each; Table~\ref{tab:n12}). These comparisons count evaluations. In time, our implementations of the classical solvers took $1.4\times10^{-6}$ to $1.6\times10^{-5}$~s per evaluation on the main-family subproblems at $n=12$ (plus 0.016~s to solve the semidefinite program of semidefinite-programming rounding followed by local search, SDP+LS), against $3.30\times10^{-4}$~s of quantum processing time per shot, which excludes queueing and compilation, so a comparison in time would favor the classical solvers further. Because the noiseless state already needed more evaluations than SDP+LS and tabu search with restarts on these subproblems, the planted circuits address whether an advantage can survive execution.

On the planted subproblems, transferred QAOA finds the optimum in noiseless simulation in 98\% of 768-shot blocks, more often than tabu search with restarts (78\%, $P=0.027$), local search (62\%), PT, and tuned SA (54\%; $P\le0.004$ for each), but not more often than SDP+LS (82\%, $P=0.43$). On hardware without suppression, the transferred state retained 19\% of its noiseless probability of the optimum (floor-corrected; Table~\ref{tab:etaq}) and found the optimum in 62\% of blocks: as often as local search, on average less often than tabu search with restarts and SDP+LS, and more often than PT and tuned SA, but none of these differences was significant ($P$ from 0.14 to 0.97). With dynamical decoupling and twirling, it found the optimum in 40\% of blocks, less often than tabu search with restarts ($P=0.016$), with its other differences not significant ($P$ from 0.063 to 0.32) (Table~\ref{tab:n12}, Figure~\ref{fig:n12second}a). On this device and day, for ten instances of the first subproblem at $n=12$, transferred QAOA on hardware therefore no longer exceeded any of the local heuristics, which it exceeds in noiseless simulation. These comparisons are post hoc, with ten instances each.

\subsection{Suppression Settings and Drift}\label{subsec:suppression}

In the second submission, dynamical decoupling alone raised the retention over no suppression on the main family (means 0.46 against 0.44, $P=0.010$) and found the optimum in 90\% of blocks, whereas twirling, alone or combined with dynamical decoupling, lowered it on all ten main-family instances (0.38 and 0.37; $P=0.002$ for each), and the combination lowered it on all ten planted circuits (0.23 against 0.36; $P=0.002$). The lower retention of the combined setting in the first submission and in both 8-qubit sessions is therefore consistent with an effect of twirling rather than of dynamical decoupling, although each setting was run once per submission. Between the submissions, the mean retention without suppression drifted from 0.40 to 0.44 ($P=0.014$), a change comparable in size to the effects of suppression. Readout mitigation changed the median probability of the optimum by at most $3\times10^{-4}$ in every mode. These exploratory comparisons describe this device on these days.

\subsection{Hardware Inside ADMM}\label{subsec:admm}

Two small studies ran six ADMM iterations of problem \eqref{eq:mbo} with the binary step executed on the device (768 shots per iteration, best sample of each iteration). In the first, on three main-family instances with XY4 dynamical decoupling, the hardware solved 17 of 18 subproblems and reached the same best feasible solutions as noiseless sampling, but the reference trajectories never changed their bitstring and the hardware trajectory changed only at its one missed subproblem, so the device mostly had to resample one subproblem optimum. The second study therefore selected, in noiseless simulation and before submission, the first three planted instances whose transferred-angle ADMM trajectory changes its bitstring at least twice in six iterations (instances 0, 34, and 42), and ran them without suppression. The hardware changed the bitstring two to four times per instance (noiseless sampling: two or three) but solved only 5 of its 18 subproblems exactly, against 15 for noiseless sampling and 10 for the device-matched simulation, and found the problem optimum on one of the three instances (best feasible normalized gap 0.017 and 0.044 on the others), whereas noiseless sampling, the device-matched simulation, and an exact solver found it on all three. With three instances these are illustrations rather than measurements: when the ADMM trajectory depends on the subproblem solutions, hardware errors change the decomposition's outcome.

\section{Discussion}\label{sec:discussion}

\emph{Validation practice.} A mean-gap or mean-cost comparison between a device and a model answers how far the output is from uniform, which Proposition~\ref{prop:depol} shows is the right quantity only under global depolarization, and the threshold-resolved retention shows is dominated by the avoidance of poor bitstrings. For circuits meant to sample optimal or near-optimal solutions, a validation should report the retention of the optimum and of near-optimal sets, floor-corrected, alongside the mean, with the number of optimal shots on which it rests. For control circuits with near-uniform output, the mean-gap retention is ill-conditioned and should not be used to compare devices or models. A device-matched noise model that predicts the mean gap well can still overestimate the retention of the optimum several-fold, as here. Because the effective fidelity can depend on the angles, a validation should also use the angles of the production circuits rather than random ones.

\emph{What the device preserves.} The retention of the optimum was three to eight times lower than the mean-gap retention on the main family and about twice lower, on fewer optimal shots, on the planted family. The effective fidelity also depended on the angles: without suppression, random-angle circuits with the same gate count and a well-determined $\eta$ retained between $-0.09$ and 0.81 of their mean-gap improvement, where the transferred-angle circuits retained 0.28 to 0.48. Because near-optimal sets were lost about as much as the optimum and the optimum's single-flip neighbors did not absorb the loss, the effect is not explained by a few bit flips moving the state to nearby solutions. A device-matched model built from reported gate, readout, and relaxation errors reproduces the ordering $\eta_q<\eta$ at a higher level, which suggests that local noise already makes concentrated probability more fragile than the mean, while the remaining gap between model and device is large. Identifying the responsible error processes, for example crosstalk or coherent errors that the snapshot model omits, was outside the scope of this study.

\emph{Consequences for optimization.} At these sizes, the time to solution on hardware was 5 to 14 times the noiseless value, and the planted-family advantage over local heuristics that transferred QAOA shows in noiseless simulation did not survive. Estimates of hardware time to solution from mean-cost retention would overstate the retained advantage by the factor $\eta/\eta_q$. Suppression settings matter at the level of the observed effects: on this device, twirling lowered and dynamical decoupling raised the retention, and drift between submissions was of comparable size, so single executions of each setting should be read as descriptions of the device on a given day.

\emph{Limitations.} The study uses one device, five instances at $n=8$, ten instances per family at $n=12$, depth $p=2$, and the first ADMM subproblem only, with each suppression setting run once per submission, and gate errors and coherence times at execution time were not recorded. The probability of the optimum rests on few optimal shots per instance, so per-instance $\eta_q$ is noisy, although the ordering $\eta_q<\eta$ is consistent across runs. The floor correction assumes that noise moves probability toward the uniform distribution; a check of the counts found no bias toward $|0\cdots0\rangle$ on the main family but a clear one on the planted circuits without suppression (Section~\ref{subsec:threshold}). Threshold-resolved retention locates where probability was lost, not which errors caused the loss. All hardware tests are exploratory.

\section{Conclusion}\label{sec:conclusions}

On a superconducting processor, depth-2 QAOA circuits with transferred angles retained 29\% to 46\% of their noiseless mean-gap improvement at 12 qubits but only 5\% to 15\% of their probability of the optimum, contrary to the global-depolarization prediction that both should be equal. The loss extends to near-optimal solutions, whereas the avoidance of poor solutions is preserved best, so mean-gap agreement between device and model mostly certifies the latter. On the first submission, a device-matched noise model reproduced the pattern but overestimated the retention of the optimum five- to eightfold, and the hardware time to solution exceeded that of tuned classical heuristics. On a rugged planted family, the advantage of transferred QAOA over local heuristics in noiseless simulation did not survive execution. Hardware validations of optimization circuits should report the floor-corrected retention of the optimum and of near-optimal sets, and studies of suppression should repeat each setting to separate its effect from drift.

\section{Experimental Section}\label{sec:methods}

\subsection{Device, Circuits, and Submissions}\label{subsec:device}

All executions used \texttt{ibm\_marrakesh}, an IBM Quantum Heron r2 processor with 156 qubits, through the Qiskit Runtime sampler, with circuits built in Qiskit 2.5.2\cite{javadi2024quantum} and transpiled at optimization level 3 for the device's coupling map and native gates with a fixed transpiler seed. Each circuit measured all qubits with 4096 shots. The two 8-qubit sessions were submitted on 28 September 2026 at about 07:24 and 17:12 UTC, the first 12-qubit submission and the first ADMM study on 30 September 2026 at about 14:14 and 15:35 UTC, and the second 12-qubit submission and the second ADMM study on 1 October 2026 at 14:20 and 14:26 UTC; the calibration time of the device was recorded only for the second 12-qubit submission (13:34 UTC). For the executed qubits, the readout assignment errors measured with the calibration circuits described below had medians of 0.0095 (0 read as 1) and 0.0117 (1 read as 0) in the first 12-qubit submission and of 0.0021 and 0.0115 in the second; the largest error on any executed qubit was 0.061 in the first and 0.025 in the second. The first submission used 15 physical qubits because two of its 20 circuits were routed to a second layout that shares nine qubits with the layout of the others. Readout errors were not measured in the 8-qubit sessions, and other device properties at execution time, such as gate errors and coherence times, were not recorded with the jobs.

\emph{8-qubit validation.} Two sessions, submitted about ten hours apart from two complete executions of our code, ran the first subproblems of five main-family instances at $n=8$ with random and with transferred angles, each with and without suppression. Each transpiled circuit contains 188 two-qubit gates, against 112 in the unrouted logical circuit.

\emph{12-qubit submissions.} The first submission ran the first subproblems of ten main-family instances at $n=12$ with transferred and with random angles, with and without suppression. The second submission ran the same transferred-angle circuits without suppression, with dynamical decoupling alone, with twirling alone, and with both, and the ten planted-family circuits without suppression and with both, on one fixed layout of 12 physical qubits. Every routed 12-qubit circuit has 526 two-qubit gates (264 in the logical circuit) at a depth of 736 to 756. Qiskit Runtime reported 54~s of quantum processing time for the $163\,840$ shots of the first submission and 73~s for the $204\,800$ shots of the five jobs of the second submission that reported usage (its sixth reported none), that is, $3.30\times10^{-4}$ and $3.56\times10^{-4}$~s per shot.

\emph{Suppression and mitigation.} Suppression combines XY4 dynamical decoupling\cite{ezzell2023dynamical} with gate and measurement twirling.\cite{wallman2016noise} All other sampler options were left at their documented defaults: dynamical decoupling used as-late-as-possible scheduling with extra slack in the middle of each sequence, and twirling used the default strategy with automatic randomization, which for 4096 shots gives 32 randomizations of 128 shots. Readout mitigation,\cite{bravyi2021mitigating} with a tensored assignment model calibrated on the executed qubits by preparing all zeros and all ones (4096 shots each), was applied in a secondary analysis only; all reported values use unmitigated counts unless stated otherwise.

\emph{Noise models.} The device-matched simulation runs the transpiled circuit on a simulator configured from a snapshot of the device's reported gate, readout, and relaxation errors (Qiskit Aer with the \texttt{FakeMarrakesh} backend); at $n=8$ it draws 8192 shots, and a second model applies the same noise to the unrouted logical circuit. At $n=12$ the device-matched simulation of the first submission draws 4096 shots per circuit. Its retention of the optimum was not recorded when it was first run; we re-executed it with the recorded transpiler and simulator seeds, which reproduced its recorded $\eta$ of every instance exactly.

\subsection{Scoring and Classical Comparators}\label{subsec:scoring}

A sampler solves a subproblem in a block of $\mathcal{B}$ evaluations if the block contains an optimal bitstring. We split the 4096 shots of each circuit into five consecutive blocks of 768 shots and report the fraction of blocks containing the optimum, against five independent runs of 768 evaluations for each classical solver. One evaluation is one candidate bitstring whose cost the solver obtains: a shot, a Metropolis proposal of simulated annealing (SA), a random draw, or a hyperplane rounding; tabu search and local search are charged $n$ evaluations per scan of the $n$ one-flip neighbors. The comparators are SA\cite{kirkpatrick1983optimization} with geometric cooling from temperature 2 to 0.01, SA tuned on a grid of 36 configurations (initial temperature 1, final temperature 0.1, four restarts), and random search for the first submission; and, for the planted family, SA tuned on a grid of 125 configurations (0.5, 0.005, 16 restarts), tabu search with restarts\cite{glover1989tabu,palubeckis2004multistart} (tenure 9, restart after 12 non-improving steps), PT\cite{hukushima1996exchange,earl2005parallel} (16 replicas on temperatures from 0.02 to 0.3), steepest-descent local search with random restarts, semidefinite-programming rounding followed by local search (SDP+LS), and random search. Each tuned solver was tuned per family by the mean normalized gap at 768 evaluations on the 36 held-out subproblems on which the angles were trained.

The time to solution $\mathrm{TTS}_{99}$ is the number of evaluations needed to find the optimum with probability 0.99.\cite{ronnow2014defining,albash2018demonstration} For a circuit placing probability $q$ on the optimum, $\mathrm{TTS}_{99}=\ln0.01/\ln(1-q)$, with $q$ exact for noiseless circuits and estimated from the 4096 shots on hardware. For the classical solvers on the main-family instances of the first submission (tabu search with restarts with tenure 3 and restart after 6 non-improving steps, PT with 8 replicas on temperatures from 0.005 to 3, SA with one anneal per run at the temperatures of the 125-configuration grid, local search, and SDP+LS), the run budget $b^\star$ was selected at $n=12$ from 25 values between 8 and 32\,768 by minimizing the median $\mathrm{TTS}_{99}$ on the second ADMM subproblem of 20 held-out instances (256 runs per budget), and each subproblem was scored from 1024 runs at $b^\star$ as $b^\star\ln0.01/\ln(1-\hat p)$, where $\hat p$ is the fraction of runs that found the optimum.

\subsection{Statistics}\label{subsec:stats}

The unit of analysis is the instance. Paired comparisons use the exact two-sided Wilcoxon signed-rank test;\cite{wilcoxon1945individual} with ten instances the smallest attainable $P$ is $2^{-9}\approx0.002$. Intervals for means over instances are percentile bootstrap 95\% intervals;\cite{efron1979bootstrap} the interval of a median $\eta_q$ resamples instances and, within each, the binomial count of optimal shots. The 25 hardware tests computed by our analysis code, all exploratory, are recorded in a log: 15 have $P$ below the Bonferroni threshold $0.05/25=0.002$, and 21 remain significant after Holm correction\cite{holm1979simple} within their five families. The device-model re-execution, the binomial check of the optimal counts, the Hamming-weight check, the threshold-resolved retention, the Hamming-shell analysis, the comparisons with classical solvers on the planted family, and the effect of readout mitigation are post hoc and outside the log; the planted-family comparisons use the same signed-rank test.

\section*{Author Contributions}

\textbf{Sanim Yousuf Fahim}: conceptualization, methodology, software, investigation, formal analysis, data curation, visualization, writing -- original draft. \textbf{Mohammad Arif Hossain}: conceptualization, methodology, validation, supervision, project administration, writing -- review and editing.

\section*{Use of AI Tools}

The authors used Claude (Anthropic) Opus 5.5 model in September–October 2026 to assist with editing the text and writing post-processing scripts. The authors designed the study, ran all experiments, verified every reported number, reference, and script output against the computed results, and take full responsibility for the content. No sensitive or human-subject data were involved.

\section*{Acknowledgments}

We acknowledge the use of IBM Quantum services for this work. The views expressed are those of the authors and do not reflect the official policy or position of IBM or the IBM Quantum team. The authors received no specific funding for this work.

\section*{Conflicts of Interest}

The authors declare no conflicts of interest.

\section*{Ethics Statement}

This study involved no human participants, human data, or animals, so ethics approval and informed consent were not required. All figures and tables were created by the authors from their own data, and no material was reproduced from other sources.

\section*{Data Availability Statement}

The hardware counts, job identifiers, circuits, readout calibration, per-instance results, test log, and the code that supports the findings of this study are openly available at \url{https://github.com/sanim-yousuf/qaoa-admm-benchmark}.

\bibliographystyle{wileyNJD-Chicago}
\bibliography{references}

\begin{thebibliography}{42}
\providecommand{\natexlab}[1]{#1}
\providecommand{\url}[1]{\texttt{#1}}
\expandafter\ifx\csname urlstyle\endcsname\relax
  \providecommand{\doi}[1]{doi: #1}\else
  \providecommand{\doi}{doi: \begingroup \urlstyle{rm}\Url}\fi

\bibitem[\protect\citeauthoryear{Farhi, Goldstone, and Gutmann}{Farhi
  et~al.}{}]{farhi2014quantum}
Edward Farhi, Jeffrey Goldstone, and Sam Gutmann. 2014.
\newblock ``A Quantum Approximate Optimization Algorithm.''   arXiv:1411.4028.

\bibitem[\protect\citeauthoryear{Harrigan, Sung, Neeley, et~al.}{Harrigan
  et~al.}{}]{harrigan2021quantum}
Matthew~P. Harrigan, Kevin~J. Sung, Matthew Neeley, et~al.
\newblock
\newblock ``Quantum Approximate Optimization of Non-Planar Graph Problems on a
  Planar Superconducting Processor.''  {\it Nature Physics\/}~17, no. 3 (2021):
  332--336.
\newblock \url{https://doi.org/10.1038/s41567-020-01105-y}

\bibitem[\protect\citeauthoryear{Weidenfeller, Valor, Gacon,
  et~al.}{Weidenfeller et~al.}{}]{weidenfeller2022scaling}
Johannes Weidenfeller, Lucia~C. Valor, Julien Gacon, et~al.
\newblock
\newblock ``Scaling of the Quantum Approximate Optimization Algorithm on
  Superconducting Qubit Based Hardware.''  {\it Quantum\/}~6 (2022): 870.
\newblock \url{https://doi.org/10.22331/q-2022-12-07-870}

\bibitem[\protect\citeauthoryear{Pelofske, B{\"a}rtschi, Cincio, Golden, and
  Eidenbenz}{Pelofske et~al.}{}]{pelofske2024scaling}
Elijah Pelofske, Andreas B{\"a}rtschi, Lukasz Cincio, John Golden, and Stephan
  Eidenbenz.
\newblock
\newblock ``Scaling Whole-Chip {QAOA} for Higher-Order {Ising} Spin Glass
  Models on Heavy-Hex Graphs.''  {\it npj Quantum Information\/}~10, no. 1
  (2024): 109.
\newblock \url{https://doi.org/10.1038/s41534-024-00906-w}

\bibitem[\protect\citeauthoryear{Pagano, Bapat, Becker, et~al.}{Pagano
  et~al.}{}]{pagano2020quantum}
Guido Pagano, Aniruddha Bapat, Patrick Becker, et~al.
\newblock
\newblock ``Quantum Approximate Optimization of the Long-Range {Ising} Model
  with a Trapped-Ion Quantum Simulator.''  {\it Proceedings of the National
  Academy of Sciences of the United States of America\/}~117, no. 41 (2020):
  {25396}--{25401}.
\newblock \url{https://doi.org/10.1073/pnas.2006373117}

\bibitem[\protect\citeauthoryear{Preskill}{Preskill}{}]{preskill2018nisq}
John Preskill.
\newblock
\newblock ``Quantum Computing in the {NISQ} Era and Beyond.''  {\it
  Quantum\/}~2 (2018): 79.
\newblock \url{https://doi.org/10.22331/q-2018-08-06-79}

\bibitem[\protect\citeauthoryear{Stilck~França and
  García-Patrón}{Stilck~França and
  García-Patrón}{}]{stilckfranca2021limitations}
Daniel Stilck~França and Raúl García-Patrón.
\newblock
\newblock ``Limitations of Optimization Algorithms on Noisy Quantum Devices.''
  {\it Nature Physics\/}~17 (2021): 1221--1227.
\newblock \url{https://doi.org/10.1038/s41567-021-01356-3}

\bibitem[\protect\citeauthoryear{Lubinski, Coffrin, McGeoch, et~al.}{Lubinski
  et~al.}{}]{lubinski2024optimization}
Thomas Lubinski, Carleton Coffrin, Catherine McGeoch, et~al.
\newblock
\newblock ``Optimization Applications as Quantum Performance Benchmarks.''
  {\it ACM Transactions on Quantum Computing\/}~5, no. 3 (2024): 1--44.
\newblock Art. no. 18.
\newblock \url{https://doi.org/10.1145/3678184}

\bibitem[\protect\citeauthoryear{Abbas, Ambainis, Augustino, et~al.}{Abbas
  et~al.}{}]{abbas2024challenges}
Amira Abbas, Andris Ambainis, Brandon Augustino, et~al.
\newblock
\newblock ``Challenges and Opportunities in Quantum Optimization.''  {\it
  Nature Reviews Physics\/}~6, no. 12 (2024): 718--735.
\newblock \url{https://doi.org/10.1038/s42254-024-00770-9}

\bibitem[\protect\citeauthoryear{Koch, Bernal~Neira, Chen, et~al.}{Koch
  et~al.}{}]{koch2026quantum}
Thorsten Koch, David~E. Bernal~Neira, Ying Chen, et~al.
\newblock
\newblock ``The Quantum Optimization Benchmarking Library.''  {\it Nature
  Computational Science\/}~6, no. 6 (2026): 653--671.
\newblock \url{https://doi.org/10.1038/s43588-026-00991-1}

\bibitem[\protect\citeauthoryear{R{\o}nnow, Wang, Job, et~al.}{R{\o}nnow
  et~al.}{}]{ronnow2014defining}
Troels~F. R{\o}nnow, Zhihui Wang, Joshua Job, et~al.
\newblock
\newblock ``Defining and Detecting Quantum Speedup.''  {\it Science\/}~345, no.
  6195 (2014): 420--424.
\newblock \url{https://doi.org/10.1126/science.1252319}

\bibitem[\protect\citeauthoryear{Albash and Lidar}{Albash and
  Lidar}{}]{albash2018demonstration}
Tameem Albash and Daniel~A. Lidar.
\newblock
\newblock ``Demonstration of a Scaling Advantage for a Quantum Annealer over
  Simulated Annealing.''  {\it Physical Review X\/}~8, no. 3 (2018): 031016.
\newblock \url{https://doi.org/10.1103/PhysRevX.8.031016}

\bibitem[\protect\citeauthoryear{Lotshaw, Nguyen, Santana, et~al.}{Lotshaw
  et~al.}{}]{lotshaw2022scaling}
Phillip~C. Lotshaw, Thien Nguyen, Anthony Santana, et~al.
\newblock
\newblock ``Scaling Quantum Approximate Optimization on Near-Term Hardware.''
  {\it Scientific Reports\/}~12 (2022): 12388.
\newblock \url{https://doi.org/10.1038/s41598-022-14767-w}

\bibitem[\protect\citeauthoryear{Boyd, Parikh, Chu, Peleato, and Eckstein}{Boyd
  et~al.}{}]{boyd2011distributed}
Stephen Boyd, Neal Parikh, Eric Chu, Borja Peleato, and Jonathan Eckstein.
\newblock
\newblock ``Distributed Optimization and Statistical Learning via the
  Alternating Direction Method of Multipliers.''  {\it Foundations and Trends
  in Machine Learning\/}~3, no. 1 (2011): 1--122.
\newblock \url{https://doi.org/10.1561/2200000016}

\bibitem[\protect\citeauthoryear{Gambella and Simonetto}{Gambella and
  Simonetto}{}]{gambella2020multiblock}
Claudio Gambella and Andrea Simonetto.
\newblock
\newblock ``Multiblock {ADMM} Heuristics for Mixed-Binary Optimization on
  Classical and Quantum Computers.''  {\it IEEE Transactions on Quantum
  Engineering\/}~1 (2020): 1--22.
\newblock Art. no. 3102022.
\newblock \url{https://doi.org/10.1109/TQE.2020.3033139}

\bibitem[\protect\citeauthoryear{Brand{\~a}o, Broughton, Farhi, Gutmann, and
  Neven}{Brand{\~a}o et~al.}{}]{brandao2018forfixed}
Fernando G. S.~L. Brand{\~a}o, Michael Broughton, Edward Farhi, Sam Gutmann,
  and Hartmut Neven. 2018.
\newblock ``For Fixed Control Parameters the Quantum Approximate Optimization
  Algorithm's Objective Function Value Concentrates for Typical Instances.''
  arXiv:1812.04170.

\bibitem[\protect\citeauthoryear{Akshay, Rabinovich, Campos, and
  Biamonte}{Akshay et~al.}{}]{akshay2021concentration}
V.~Akshay, D.~Rabinovich, E.~Campos, and J.~Biamonte.
\newblock
\newblock ``Parameter Concentrations in Quantum Approximate Optimization.''
  {\it Physical Review A\/}~104, no. 1 (2021): L010401.
\newblock \url{https://doi.org/10.1103/PhysRevA.104.L010401}

\bibitem[\protect\citeauthoryear{Shaydulin, Lotshaw, Larson, Ostrowski, and
  Humble}{Shaydulin et~al.}{}]{shaydulin2023parameter}
Ruslan Shaydulin, Phillip~C. Lotshaw, Jeffrey Larson, James Ostrowski, and
  Travis~S. Humble.
\newblock
\newblock ``Parameter Transfer for Quantum Approximate Optimization of Weighted
  {MaxCut}.''  {\it ACM Transactions on Quantum Computing\/}~4, no. 3 (2023):
  1--15.
\newblock Art. no. 19.
\newblock \url{https://doi.org/10.1145/3584706}

\bibitem[\protect\citeauthoryear{Monta{\~n}ez-Barrera and
  Michielsen}{Monta{\~n}ez-Barrera and
  Michielsen}{}]{montanezbarrera2025toward}
J.~A. Monta{\~n}ez-Barrera and Kristel Michielsen.
\newblock
\newblock ``Toward a Linear-Ramp {QAOA} Protocol: Evidence of a Scaling
  Advantage in Solving Some Combinatorial Optimization Problems.''  {\it npj
  Quantum Information\/}~11 (2025): 131.
\newblock \url{https://doi.org/10.1038/s41534-025-01082-1}

\bibitem[\protect\citeauthoryear{Xue, Chen, Wu, and Guo}{Xue
  et~al.}{}]{xue2021effects}
Cheng Xue, Zhao-Yun Chen, Yu-Chun Wu, and Guo-Ping Guo.
\newblock
\newblock ``Effects of Quantum Noise on Quantum Approximate Optimization
  Algorithm.''  {\it Chinese Physics Letters\/}~38, no. 3 (2021): 030302.
\newblock \url{https://doi.org/10.1088/0256-307X/38/3/030302}

\bibitem[\protect\citeauthoryear{Marshall, Wudarski, Hadfield, and
  Hogg}{Marshall et~al.}{}]{marshall2020characterizing}
Jeffrey Marshall, Filip Wudarski, Stuart Hadfield, and Tad Hogg.
\newblock
\newblock ``Characterizing Local Noise in {QAOA} Circuits.''  {\it IOP
  SciNotes\/}~1, no. 2 (2020): 025208.
\newblock \url{https://doi.org/10.1088/2633-1357/abb0d7}

\bibitem[\protect\citeauthoryear{Wang, Fontana, Cerezo, et~al.}{Wang
  et~al.}{}]{wang2021noise}
Samson Wang, Enrico Fontana, M.~Cerezo, et~al.
\newblock
\newblock ``Noise-Induced Barren Plateaus in Variational Quantum Algorithms.''
  {\it Nature Communications\/}~12 (2021): 6961.
\newblock \url{https://doi.org/10.1038/s41467-021-27045-6}

\bibitem[\protect\citeauthoryear{Arute, Arya, Babbush, et~al.}{Arute
  et~al.}{}]{arute2019quantum}
Frank Arute, Kunal Arya, Ryan Babbush, et~al.
\newblock
\newblock ``Quantum Supremacy Using a Programmable Superconducting Processor.''
   {\it Nature\/}~574, no. 7779 (2019): 505--510.
\newblock \url{https://doi.org/10.1038/s41586-019-1666-5}

\bibitem[\protect\citeauthoryear{Dalzell, Hunter-Jones, and
  Brand{\~a}o}{Dalzell et~al.}{}]{dalzell2021random}
Alexander~M. Dalzell, Nicholas Hunter-Jones, and Fernando G. S.~L. Brand{\~a}o.
  2021.
\newblock ``Random Quantum Circuits Transform Local Noise into Global White
  Noise.''   arXiv:2111.14907.

\bibitem[\protect\citeauthoryear{Ezzell, Pokharel, Tewala, Quiroz, and
  Lidar}{Ezzell et~al.}{}]{ezzell2023dynamical}
Nic Ezzell, Bibek Pokharel, Lina Tewala, Gregory Quiroz, and Daniel~A. Lidar.
\newblock
\newblock ``Dynamical Decoupling for Superconducting Qubits: A Performance
  Survey.''  {\it Physical Review Applied\/}~20, no. 6 (2023): 064027.
\newblock \url{https://doi.org/10.1103/PhysRevApplied.20.064027}

\bibitem[\protect\citeauthoryear{Wallman and Emerson}{Wallman and
  Emerson}{}]{wallman2016noise}
Joel~J. Wallman and Joseph Emerson.
\newblock
\newblock ``Noise Tailoring for Scalable Quantum Computation via Randomized
  Compiling.''  {\it Physical Review A\/}~94, no. 5 (2016): 052325.
\newblock \url{https://doi.org/10.1103/PhysRevA.94.052325}

\bibitem[\protect\citeauthoryear{Temme, Bravyi, and Gambetta}{Temme
  et~al.}{}]{temme2017error}
Kristan Temme, Sergey Bravyi, and Jay~M. Gambetta.
\newblock
\newblock ``Error Mitigation for Short-Depth Quantum Circuits.''  {\it Physical
  Review Letters\/}~119, no. 18 (2017): 180509.
\newblock \url{https://doi.org/10.1103/PhysRevLett.119.180509}

\bibitem[\protect\citeauthoryear{Kandala, Temme, C{\'o}rcoles, Mezzacapo, Chow,
  and Gambetta}{Kandala et~al.}{}]{kandala2019error}
Abhinav Kandala, Kristan Temme, Antonio~D. C{\'o}rcoles, Antonio Mezzacapo,
  Jerry~M. Chow, and Jay~M. Gambetta.
\newblock
\newblock ``Error Mitigation Extends the Computational Reach of a Noisy Quantum
  Processor.''  {\it Nature\/}~567, no. 7749 (2019): 491--495.
\newblock \url{https://doi.org/10.1038/s41586-019-1040-7}

\bibitem[\protect\citeauthoryear{Bravyi, Sheldon, Kandala, McKay, and
  Gambetta}{Bravyi et~al.}{}]{bravyi2021mitigating}
Sergey Bravyi, Sarah Sheldon, Abhinav Kandala, David~C. McKay, and Jay~M.
  Gambetta.
\newblock
\newblock ``Mitigating Measurement Errors in Multiqubit Experiments.''  {\it
  Physical Review A\/}~103, no. 4 (2021): 042605.
\newblock \url{https://doi.org/10.1103/PhysRevA.103.042605}

\bibitem[\protect\citeauthoryear{Kim, Eddins, Anand, et~al.}{Kim
  et~al.}{}]{kim2023evidence}
Youngseok Kim, Andrew Eddins, Sajant Anand, et~al.
\newblock
\newblock ``Evidence for the Utility of Quantum Computing Before Fault
  Tolerance.''  {\it Nature\/}~618, no. 7965 (2023): 500--505.
\newblock \url{https://doi.org/10.1038/s41586-023-06096-3}

\bibitem[\protect\citeauthoryear{Hamze, Raymond, Pattison, Biswas, and
  Katzgraber}{Hamze et~al.}{}]{hamze2020wishart}
Firas Hamze, Jack Raymond, Christopher~A. Pattison, Katja Biswas, and Helmut~G.
  Katzgraber.
\newblock
\newblock ``Wishart Planted Ensemble: A Tunably Rugged Pairwise {Ising} Model
  with a First-Order Phase Transition.''  {\it Physical Review E\/}~101, no. 5
  (2020): 052102.
\newblock \url{https://doi.org/10.1103/PhysRevE.101.052102}

\bibitem[\protect\citeauthoryear{Sureshbabu, Herman, Shaydulin,
  et~al.}{Sureshbabu et~al.}{}]{sureshbabu2024parameter}
Shree~Hari Sureshbabu, Dylan Herman, Ruslan Shaydulin, et~al.
\newblock
\newblock ``Parameter Setting in Quantum Approximate Optimization of Weighted
  Problems.''  {\it Quantum\/}~8 (2024): 1231.
\newblock \url{https://doi.org/10.22331/q-2024-01-18-1231}

\bibitem[\protect\citeauthoryear{Nelder and Mead}{Nelder and
  Mead}{}]{nelder1965simplex}
J.~A. Nelder and R.~Mead.
\newblock
\newblock ``A Simplex Method for Function Minimization.''  {\it The Computer
  Journal\/}~7, no. 4 (1965): 308--313.
\newblock \url{https://doi.org/10.1093/comjnl/7.4.308}

\bibitem[\protect\citeauthoryear{Javadi-Abhari, Treinish, Krsulich,
  et~al.}{Javadi-Abhari et~al.}{}]{javadi2024quantum}
Ali Javadi-Abhari, Matthew Treinish, Kevin Krsulich, et~al. 2024.
\newblock ``Quantum Computing with {Qiskit}.''   arXiv:2405.08810.

\bibitem[\protect\citeauthoryear{Kirkpatrick, Gelatt, and Vecchi}{Kirkpatrick
  et~al.}{}]{kirkpatrick1983optimization}
S.~Kirkpatrick, C.~D. Gelatt, Jr., and M.~P. Vecchi.
\newblock
\newblock ``Optimization by Simulated Annealing.''  {\it Science\/}~220, no.
  4598 (1983): 671--680.
\newblock \url{https://doi.org/10.1126/science.220.4598.671}

\bibitem[\protect\citeauthoryear{Glover}{Glover}{}]{glover1989tabu}
Fred Glover.
\newblock
\newblock ``Tabu Search---{Part I}.''  {\it ORSA Journal on Computing\/}~1, no.
  3 (1989): 190--206.
\newblock \url{https://doi.org/10.1287/ijoc.1.3.190}

\bibitem[\protect\citeauthoryear{Palubeckis}{Palubeckis}{}]{palubeckis2004multistart}
Gintaras Palubeckis.
\newblock
\newblock ``Multistart Tabu Search Strategies for the Unconstrained Binary
  Quadratic Optimization Problem.''  {\it Annals of Operations Research\/}~131,
  no. 1 (2004): 259--282.
\newblock \url{https://doi.org/10.1023/B:ANOR.0000039522.58036.68}

\bibitem[\protect\citeauthoryear{Hukushima and Nemoto}{Hukushima and
  Nemoto}{}]{hukushima1996exchange}
Koji Hukushima and Koji Nemoto.
\newblock
\newblock ``Exchange {Monte Carlo} Method and Application to Spin Glass
  Simulations.''  {\it Journal of the Physical Society of Japan\/}~65, no. 6
  (1996): 1604--1608.
\newblock \url{https://doi.org/10.1143/JPSJ.65.1604}

\bibitem[\protect\citeauthoryear{Earl and Deem}{Earl and
  Deem}{}]{earl2005parallel}
David~J. Earl and Michael~W. Deem.
\newblock
\newblock ``Parallel Tempering: Theory, Applications, and New Perspectives.''
  {\it Physical Chemistry Chemical Physics\/}~7, no. 23 (2005): 3910--3916.
\newblock \url{https://doi.org/10.1039/B509983H}

\bibitem[\protect\citeauthoryear{Wilcoxon}{Wilcoxon}{}]{wilcoxon1945individual}
Frank Wilcoxon.
\newblock
\newblock ``Individual Comparisons by Ranking Methods.''  {\it Biometrics
  Bulletin\/}~1, no. 6 (1945): 80--83.
\newblock \url{https://doi.org/10.2307/3001968}

\bibitem[\protect\citeauthoryear{Efron}{Efron}{}]{efron1979bootstrap}
Bradley Efron.
\newblock
\newblock ``Bootstrap Methods: Another Look at the Jackknife.''  {\it The
  Annals of Statistics\/}~7, no. 1 (1979): 1--26.
\newblock \url{https://doi.org/10.1214/aos/1176344552}

\bibitem[\protect\citeauthoryear{Holm}{Holm}{}]{holm1979simple}
Sture Holm.
\newblock
\newblock ``A Simple Sequentially Rejective Multiple Test Procedure.''  {\it
  Scandinavian Journal of Statistics\/}~6, no. 2 (1979): 65--70.

\end{thebibliography}

\end{document}